\documentclass[%
preprint,
 amsmath,amssymb,
 aps,
]{revtex4-1}

\usepackage{graphicx}
\usepackage{dcolumn}
\usepackage{bm}

\begin{document}


\title{Upper Bound of Collective Attacks on Quantum Key Distribution}

\author{Wei Li$^{1,2,3}$}
\author{Shengmei Zhao$^{1,2}$}%
 \email{zhaosm@njupt.edu.cn}
\affiliation{$^{1}$Nanjing University of Posts and Telecommunications, Institute of Signal Processing and Transmission, Nanjing, 210003, China.}%
\affiliation{$^{2}$Nanjing University of Posts and Telecommunications, Key Lab Broadband Wireless Communication and Sensor Network, Ministy of Education, Nanjing, 210003, China.}%
\affiliation{$^{3}$National Laboratory of Solid State Microstructures, Nanjing University, Nanjing 210093, China.}%


\date{\today}

\begin{abstract}

Evaluating the theoretical limit of the amount of information Eve can steal from a quantum key distribution protocol under given conditions is one of the most important things that need to be done in security proof. In addition to source loopholes and detection loopholes, channel attacks are considered to be the main ways of information leakage, while collective attacks are considered to be the most powerful active channel attacks. Here we deduce in detail the capability limit of Eve's collective attack, the most powerful attack scheme in channel attacks, in non-entangled quantum key distribution, like BB84 and measurement-device-independent protocols, and entangled quantum key distribution, like device-independent protocol, in which collective attack is composed of quantum weak measurement and quantum unambiguous state discrimination detection. The theoretical results show that collective attacks are equivalent in entangled and non-entangled quantum key distribution protocols. We also find that compared with the loose bond on Eve's attack provided by security proof based on entanglement purification, the security proof against collective attacks not only improves the system's tolerable bit error rate, but also improves the key rate.

\end{abstract}

\pacs{Valid PACS appear here}
\maketitle

\section{Introduction}

\par Security proof is one of the most important components of quantum key distribution (QKD)\cite{lo1999unconditional,shor2000simple,mayers2001unconditional,gottesman2004security,pirandola2019advances}. It is not only used to evaluate the validity of a protocol, but also to give the maximum key rate that the protocol can extract. A QKD protocol is considered to be secure in this case that after the two communication parties, Alice and Bob, complete the transmission and measurement of quantum states and all post-processing, the probability of eavesdropper Eve acquiring their shared key tends to be exponentially infinitesimal. The security of QKD can be divided into two categories. The first category belongs to security loopholes, such as imperfect light source\cite{xu2010experimental,tang2013source,gisin2006trojan-horse,jain2014trojan-horse,brassard2000limitations,lutkenhaus2000security} and detection loopholes\cite{lutkenhaus1999estimates,lutkenhaus2000security,gerhardt2011full-field,makarov2006effects,qi2007time-shift,lydersen2011controlling,lydersen2010hacking}. Under appropriate attack schemes, the information of shared keys between Alice and Bob can be eavesdropped almost entirely. At present, many approaches have been proposed to close these loopholes, such as decoy-state protocol for imperfect light source\cite{hwang2003quantum,wang2005beating,lo2005decoy}, measurement-device-independent (MDI)-QKD protocols for detection loopholes\cite{lo2012measurement-device-independent,tamaki2012phase,ma2012alternative,braunstein2012side-channel-free}.

\par The other category of QKD security is the channel-side leakage of information introduced by imperfect devices in the transmission and measurement of quantum states, such as the beam-splitting (BS) attack caused by the attenuation of quantum states\cite{ma2018phase-matching}, and the collective attack scheme introduced by finite bit error rate (BER)\cite{biham1997security,biham2002security,acin2007device-independent}. The amount of information lost by channel-side attacks will determine the amount of information needed for secret amplification and the maximum tolerable BER by QKD. BS attack is a passive attack scheme in which the attenuated photon states can be considered to be intercepted and stored by Eve. The amount of information eavesdropped by Eve is determined by the joint measurement probability between Eve and Alice and Bob. In contrast with BS attack, collective attack is an active attack scheme. Eve uses weak measurement technology to obtain as much information as possible about the shared quantum states between Alice and Bob under certain perturbations. The security proof based on entanglement purification has given the general loose bound of information consumed by secret amplification. Although the security against collective attacks in the BB84 and B92 protocols\cite{biham1997security,biham2002security} has been proved, there is no corresponding expression of the general bound. Furthermore, the Holove quantity provided by DI-QKD security proof against collective attacks is not improved compared with the entanglement purification protocol\cite{acin2007device-independent}. However, after knowing Eve's attack strategy, a tighter bound for information leakage should be obtained. We already know that QKD protocols based on single-photon and entangled states have equivalence. Like entanglement purification protocols, we need to give the general expression of Holove quantity between Eve and Alice and Bob under the collective attack for all QKDs.

\par In this paper, we will evaluate the performance of Eve's collective attacks on different kinds of QKDs by deducing the Holove quantity and the tolerable maximum bit error rate. Firstly, we derive the maximum amount of information that Eve can steal using collective attack in non-entangled QKD. Eve's collective attack consists of quantum weak measurement and quantum unambiguous state discrimination detection. Eve can perform any quantum operation, and any operation of Eve only needs to obey quantum mechanics. Next, we extend Eve's collective attack to entangled QKD systems, and give the generality of collective attack in these two kinds of QKD systems.

\section{Collective attack in non-entangled QKD protocols}

\par Collective attack is currently considered to be the most powerful channel attack scheme. In collective attack, Eve attaches his quantum state to each of the states transmitted between Alice and Bob independently. Eve performs a weak measurement operation on the joint quantum states to entangle his state with the joint system of Alice and Bob. Eve then stores his quantum state and re-sends the quantum state transmitted between Alice and Bob. After Alice and Bob have completed state measurement and announcement of measurement base, Eve then makes unambiguous state discrimination (USD) measurements on the corresponding quantum state stored by him to obtain the shared information between Alice and Bob. We first show collective attacks in non-entangled two-dimensional QKD protocols, such as BB84- and MDI-QKDs. These conclusions are then extended to the entanglement-based QKD protocol.

\subsection{Weak measurement}

\begin{figure}
    \centering
    \includegraphics[width=10cm]{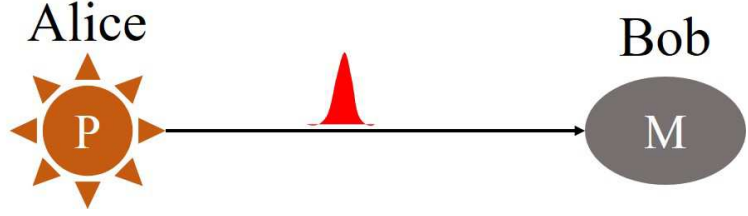}
    \caption{Schematic illustration of Eve's collective attack on BB84-QKD. Alice randomly prepares a quantum state from two mutually unbiased bases and sends it to Bob. Bob randomly chooses the measurement basis to measure quantum states. Eve uses weak measurements to implement collective attack on the quantum states transmitted in the channel. The bit error rate caused by this attack cannot exceed the agreed value of Alice and Bob.}
    \label{fig:my_label}
\end{figure}

\par Assume that the quantum state space in QKD protocol is formed by orthogonal basis vectors $\left | p \right \rangle$ and $\left | q \right \rangle$ which can interact with Eve's system without any crossing, and the bases in $Z$ space and $X$ space can be written as
\begin{equation}
\begin{split}
\left | 0 \right \rangle &=a \left | p \right \rangle + b\left | q \right \rangle, \left | 0 \right \rangle=b\left | p \right \rangle -a\left | q \right \rangle, \\
\left | + \right \rangle &=\frac{\sqrt{2}}{2} \left [ \left ( a+b \right ) \left | p \right \rangle - \left ( a-b \right )\left | q \right \rangle \right ], \left | + \right \rangle =\frac{\sqrt{2}}{2} \left [ \left ( a-b \right ) \left | p \right \rangle + \left ( a+b \right )\left | q \right \rangle \right ],
\end{split}
\end{equation}
where $a$ and $b$ are the normalization constants that satisfy $a^{2}+b^{2}=1$. In BB84, as shown in Fig. 1, Alice prepares these two sets of mutually-unbiased-bases (MUBs) with equal probabilities and sends them to Bob. Eve has a powerful attack capability in the collective attack scheme, any operation by Eve should only be restricted to the principle of quantum mechanics. Eve is free to change the values of $a$ and $b$ in Eqs. (1). Assume that Eve's initial quantum state is $\left | E \right \rangle$, the weak measurement on the joint system is represented by a unitary operator $U$,
\begin{equation}
\begin{split}
U \left | p \right \rangle \left | E \right \rangle&= \left | \alpha_{p} \right \rangle \left | E_{p} \right \rangle,\\
U \left | q \right \rangle \left | E \right \rangle&= \left | \beta_{q} \right \rangle \left | E_{q} \right \rangle,
\end{split}
\end{equation}
where $\left | \alpha_{p} \right \rangle$ and $\left | \beta_{q} \right \rangle$ are obtained by rotating $\alpha$ and $\beta$ degrees of $\left | p \right \rangle$ and $\left | q \right \rangle$ in a two-dimensional plane under weak measurements, respectively, $\left | E_{p} \right \rangle$ and $\left | E_{q} \right \rangle$ are the corresponding states stored by Eve, all the phase terms generated by the interaction are included in $\left | E_{p} \right \rangle$ and $\left | E_{q} \right \rangle$. The unitary of operator $U$ leads to the following equality
\begin{equation}
\left \langle p | q \right \rangle \left \langle E | E \right \rangle=0=\left \langle E \right | \left \langle q \right | U^{+} U \left | q \right \rangle \left | E \right \rangle=\left \langle \alpha_{p} | \beta_{q} \right \rangle \left \langle E_{p} | E_{q} \right \rangle.
\end{equation}
In general, $\left | E_{p} \right \rangle$ and $\left | E_{q} \right \rangle$ are not necessary to be orthogonal to each other. Thus for any quantum states $\left | \alpha_{p} \right \rangle$ and $\left | \beta_{q} \right \rangle$, we have $\left \langle \alpha_{p} | \beta_{q} \right \rangle=0$, which means that $\left | p \right \rangle$ and $\left | q \right \rangle$ are rotated at the same angle under weak measurement. After the unitary operation, Eve can reverse-rotate $\left | \alpha_{p} \right \rangle$ and $\left | \beta_{q} \right \rangle$ at the same angle to reduce the perturbation,
\begin{equation}
\begin{split}
T \left | p \right \rangle \left | E \right \rangle &= RU \left | p \right \rangle \left | E \right \rangle= \left | p \right \rangle \left | E_{p} \right \rangle,\\
T \left | q \right \rangle \left | E \right \rangle &= RU \left | q \right \rangle \left | E \right \rangle= \left | q \right \rangle \left | E_{q} \right \rangle.
\end{split}
\end{equation}
Then the weak measurement on the states in $Z$ bases can be written as
\begin{equation}
\begin{split}
T \left | 0 \right \rangle \left | E \right \rangle= \sqrt{p_{0,0}}\left | 0 \right \rangle \left | E_{0,0} \right \rangle+\sqrt{p_{0,1}}\left | 1 \right \rangle \left | E_{0,1} \right \rangle,\\
T \left | 1 \right \rangle \left | E \right \rangle= \sqrt{p_{1,0}}\left | 0 \right \rangle \left | E_{1,0} \right \rangle+\sqrt{p_{1,1}}\left | 1 \right \rangle \left | E_{1,1} \right \rangle,
\end{split}
\end{equation}
where $\left | E_{0,0} \right \rangle$, $\left | E_{0,1} \right \rangle$, $\left | E_{1,0} \right \rangle$ and $\left | E_{1,1} \right \rangle$ are the normalized states stored by Eve, and their expressions are
 \begin{equation}
 \begin{split}
 \left | E_{0,0} \right \rangle&=\frac{a^{2} \left | E_{p} \right \rangle+b^{2} \left | E_{q} \right \rangle}{\sqrt{1-2a^{2}b^{2}\left ( 1-\left | \left \langle E_{p} | E_{q} \right \rangle \right | \cos \theta \right )}},\\
 \left | E_{1,1} \right \rangle&=\frac{b^{2} \left | E_{p} \right \rangle+a^{2} \left | E_{q} \right \rangle}{\sqrt{1-2a^{2}b^{2}\left ( 1-\left | \left \langle E_{p} | E_{q} \right \rangle \right | \cos \theta \right )}},\\
 \left | E_{0,1} \right \rangle&=\left | E_{1,0} \right \rangle=\frac{ab\left( \left | E_{p} \right \rangle-\left | E_{q} \right \rangle \right )}{\sqrt{2a^{2}b^{2}\left ( 1-\left | \left \langle E_{p} | E_{q} \right \rangle \right | \cos \theta \right )}},
 \end{split}
 \end{equation}
where $\theta$ is the phase difference between $\left | E_{p} \right \rangle$ and $\left | E_{q} \right \rangle$. The corresponding transition probabilities $p_{0,0}$, $p_{0,1}$, $p_{1,0}$ and $p_{1,1}$ are
\begin{equation}
\begin{split}
p_{0,0}&=p_{1,1}=1-2a^{2}b^{2}\left ( 1-\left | \left \langle E_{p} | E_{q} \right \rangle \right | \cos \theta \right ),\\
p_{0,1}&=p_{1,0}=2a^{2}b^{2}\left ( 1-\left | \left \langle E_{p} | E_{q} \right \rangle \right | \cos \theta \right ).
\end{split}
\end{equation}
The weak measurement on the states in $Z$ bases can be written as
\begin{equation}
\begin{split}
T \left | + \right \rangle \left | E \right \rangle= \sqrt{p_{+,+}}\left | + \right \rangle \left | E_{+,+} \right \rangle+\sqrt{p_{+,-}}\left | - \right \rangle \left | E_{+,-} \right \rangle,\\
T \left | - \right \rangle \left | E \right \rangle= \sqrt{p_{-,+}}\left | + \right \rangle \left | E_{-,+} \right \rangle+\sqrt{p_{-,-}}\left | - \right \rangle \left | E_{-,-} \right \rangle,
\end{split}
\end{equation}
where $\left | E_{+,+} \right \rangle$, $\left | E_{+,-} \right \rangle$, $\left | E_{-,+} \right \rangle$ and $\left | E_{+,+} \right \rangle$ are the normalized states stored by Eve, and their expressions are
\begin{equation}
\begin{split}
\left | E_{+,+} \right \rangle &=\frac{\left ( a+b \right )^{2} \left | E_{p} \right \rangle+\left ( a-b \right )^{2} \left | E_{q} \right \rangle}{\sqrt{1-\frac{1}{2}\left ( 1-4a^{2}b^{2} \right ) \left ( 1-\left | \left \langle E_{p} | E_{q} \right \rangle \right | \cos \theta \right )}},\\
\left | E_{-,-} \right \rangle &=\frac{\left ( a-b \right )^{2} \left | E_{p} \right \rangle+\left ( a+b \right )^{2} \left | E_{q} \right \rangle}{\sqrt{1-\frac{1}{2}\left ( 1-4a^{2}b^{2} \right ) \left ( 1-\left | \left \langle E_{p} | E_{q} \right \rangle \right | \cos \theta \right )}},\\
\left | E_{+,-} \right \rangle &=\left | E_{-,+} \right \rangle=\frac{\left ( a^{2}-b^{2} \right )\left ( \left |  E_{q}-E_{p} \right \rangle  \right )}{\sqrt{\frac{1}{2}\left ( 1-4a^{2}b^{2} \right )\left ( 1-\left | \left \langle E_{p} | E_{q} \right \rangle \right | \cos \theta \right )}},
\end{split}
\end{equation}
and $p_{+,+}$, $p_{+,-}$, $p_{-,+}$ and $p_{-,-}$ are the corresponding transition probabilities which satisfy
\begin{equation}
\begin{split}
p_{+,+}&=p_{-,-}=1-\frac{1}{2}\left ( 1-4a^{2}b^{2} \right ) \left ( 1-\left | \left \langle E_{p} | E_{q} \right \rangle \right | \cos \theta \right ),\\
p_{+,-}&=p_{-,+}=\frac{1}{2}\left ( 1-4a^{2}b^{2} \right ) \left ( 1-\left | \left \langle E_{p} | E_{q} \right \rangle \right | \cos \theta \right ).
\end{split}
\end{equation}
From Eqs. (5) and (8), we can see that after the weak measurement performed by Eve, Eve's system is entangled with the state transmitted between Alice and Bob. From Bob's point of view, the quantum state he receives is
\begin{equation}
\rho=\frac{1}{4} [ \left ( p_{0,0} +p_{1,0} \right ) \rho_{0} +\left ( p_{0,1} +p_{1,1} \right ) \rho_{1}+\left ( p_{+,+} +p_{-,+} \right )\rho_{+}+\left ( p_{+,-} +p_{-,-} \right ) \rho_{-} ].
\end{equation}
In this case, the BER caused by weak measurement is
\begin{equation}
p_{e}=\frac{1}{4}\left ( p_{1,0}+p_{0,1}+p_{-,+} +p_{+,-} \right ),
\end{equation}
After substituting Eqs. (7) and (10) into Eqs. (12), we can get
\begin{equation}
p_{e}\geq \frac{1}{4}\left ( 1-\left | \left \langle E_{p} | E_{q} \right \rangle \right | \cos \theta \right ).
\end{equation}
It can be seen from Eqs. (13) that the BER $p_{e}$ is independent of the values of $a$ and $b$, and only depends on the inner product between $ \left | E_{p} \right \rangle$ and $ \left | E_{q} \right \rangle$. Here $\left | \left \langle E_{p} | E_{q} \right \rangle \right | $ denotes the degree of discrimination between $ \left | E_{p} \right \rangle$ and $ \left | E_{q} \right \rangle$. The smaller the value of $\left | \left \langle E_{p} | E_{q} \right \rangle \right | $, the higher the probability that Eve could distinguish $ \left | E_{p} \right \rangle$ and $ \left | E_{q} \right \rangle$. As can be seen from Eqs. (13), when the BER of the system $p_{e}$ is determined, $\theta=0$ is the best way to help Eve get information from Alice and Bob.

\subsection{Unambiguous state discrimination}

\par After Alice and Bob have announced their selected bases, Eve makes appropriate measurements on his stored quantum states to infer Alice's information. The dependence of Eve's stored states on $\left | E_{p} \right \rangle$ and $\left | E_{p} \right \rangle$ is shown in Fig. 2. Quantum states $\left | E_{0,1} \right \rangle$, $\left | E_{1,0} \right \rangle$, $\left | E_{+,-} \right \rangle$ and $\left | E_{-,+} \right \rangle$ are in the direction of $\left | E_{p} \right \rangle-\left | E_{p} \right \rangle$, $\left | E_{0,0} \right \rangle$, $\left | E_{1,1} \right \rangle$ and $\left | E_{+,-} \right \rangle$, $\left | E_{-,+} \right \rangle$ are symmetrically located on each side of $\left | E_{p} \right \rangle+\left | E_{p} \right \rangle$, respectively. There is an interesting mutually exclusive relationship between the angle of $\left | E_{0,0} \right \rangle$, $\left | E_{1,1} \right \rangle$ relative to $\left | E_{p} \right \rangle+\left | E_{p} \right \rangle$ and the angle of $\left | E_{+,+} \right \rangle$, $\left | E_{-,-} \right \rangle$ relative to $\left | E_{p} \right \rangle+\left | E_{p} \right \rangle$. For example, when $\left | E_{0,0} \right \rangle$, $\left | E_{1,1} \right \rangle$ are far from the direction of $\left | E_{p} \right \rangle+\left | E_{p} \right \rangle$, the magnitude of $\left | E_{0,1} \right \rangle$ and $\left | E_{1,0} \right \rangle$ will decrease, meanwhile $\left | E_{+,+} \right \rangle$, $\left | E_{-,-} \right \rangle$ will be near the direction of $\left | E_{p} \right \rangle+\left | E_{p} \right \rangle$, and the magnitude of $\left | E_{+,-} \right \rangle$, $\left | E_{-,+} \right \rangle$ will increase. In the extreme case, when $\left | E_{0,0} \right \rangle$, $\left | E_{1,1} \right \rangle$ are equal to $\left | E_{p} \right \rangle$ and $\left | E_{p} \right \rangle$, the magnitude of $\left | E_{0,1} \right \rangle$ and $\left | E_{1,0} \right \rangle$ is zero. At this time, $\left | E_{+,+} \right \rangle$, $\left | E_{-,-} \right \rangle$ will be strictly in the direction of $\left | E_{p} \right \rangle+\left | E_{p} \right \rangle$, and the magnitude of $\left | E_{+,-} \right \rangle$, $\left | E_{-,+} \right \rangle$ will reach the maximum value $\sqrt{\frac{1}{2}\left ( 1-\left \langle E_{p} | E_{q} \right \rangle \right )}$.

\begin{figure}
    \centering
    \includegraphics[width=8cm]{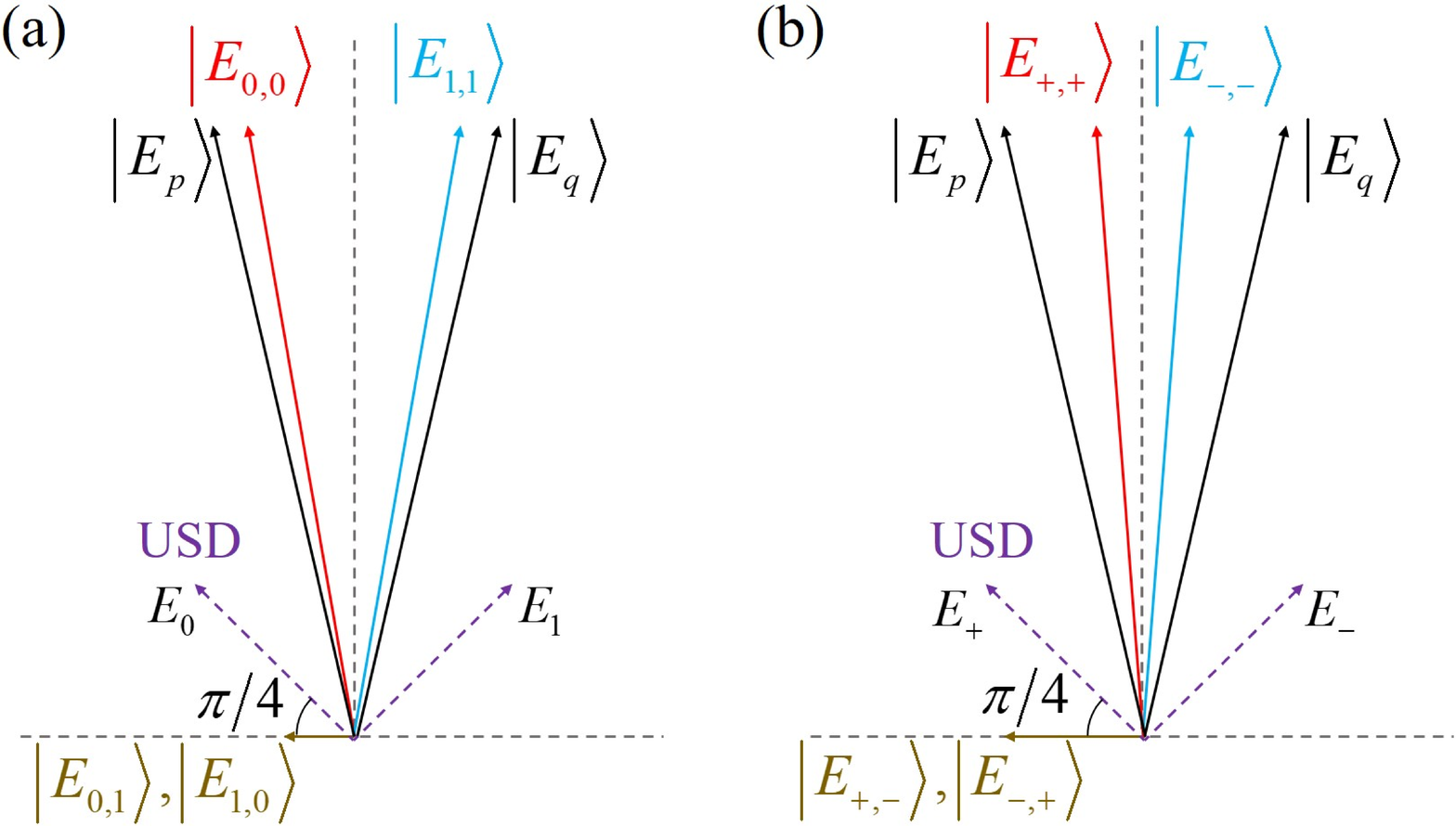}
    \caption{Schematic illustration of the quantum USD measurement of Eve's collective attack in (a) $Z$-basis and (b) $X$-basis representations. The states $\left | E_{0,0} \right \rangle$, $\left | E_{1,1} \right \rangle$ and $\left | E_{+,-} \right \rangle$, $\left | E_{-,+} \right \rangle$ are symmetrically located on each side of $\left | E_{p} \right \rangle+\left | E_{p} \right \rangle$, and $\left | E_{0,1} \right \rangle$, $\left | E_{1,0} \right \rangle$, $\left | E_{+,-} \right \rangle$ and $\left | E_{-,+} \right \rangle$ are in the direction of $\left | E_{p} \right \rangle-\left | E_{p} \right \rangle$. After Alice and Bob announce their bases publicly, Eve performs the quantum USD measurement on his stored state which is entangled with an auxiliary bit. After measuring the auxiliary bit, Eve measures his stored state along the direction of (a) $E_{0}$ and $E_{1}$ in $Z$ basis, and of (b) $E_{+}$ and $E_{-}$ in $Y$ basis.}
    \label{fig:my_label}
\end{figure}

\par Here we set the value of $a$ to $\cos \alpha$ and $b$ to $\sin \alpha$. After hearing the basis chosen by Alice and Bob, Eve should make a unambiguous discrimination between $\left | E_{0,0} \right \rangle$ and $\left | E_{1,1} \right \rangle$ or $\left | E_{+,+} \right \rangle$ and $\left | E_{-,-} \right \rangle$ to infer the state sent by Alice. The best known option for Eve is to use quantum USD measurements, which increases the detection probability from $\dfrac{1-\left \langle x | y \right \rangle^{2}}{2}$ by conventional USD measurements to $1-\left \langle x | y \right \rangle$ for non-orthogonal states $\left | x \right \rangle$, $\left | y \right \rangle$\cite{dieks1988overlap}. Suppose that at one moment, the base chosen by Alice and Bob is $Z$, Eve entangles an auxiliary bit with $\left | E_{0,0} \right \rangle$ or $\left | E_{1,1} \right \rangle$ through weak measurements\cite{ivanovic1987how,dieks1988overlap,jaeger1995optimal}. After measuring the state of the auxiliary bit, Eve chooses two orthogonal measurements $E_{0}$ and $E_{1}$ to measure the quantum states collapsed from $\left | E_{0,0} \right \rangle$ and $\left | E_{1,1} \right \rangle$, as shown in Fig. 1(a). In the plane formed by $\left | E_{p} \right \rangle$ and $\left | E_{p} \right \rangle$, the angles of the two measurements $E_{0}$ and $E_{1}$ from the direction of $\left | E_{p} \right \rangle-\left | E_{p} \right \rangle$ are $\dfrac{\pi}{4}$ and $\dfrac{3\pi}{4}$, respectively. When the value of $\alpha$ is between $0$ and $\dfrac{\pi}{4}$, USD measurements will be influenced by $\left | E_{0,1} \right \rangle$ and $\left | E_{1,0} \right \rangle$. It should be noted that in the above quantum USD measurements, $\left | E_{0,1} \right \rangle$ and $\left | E_{1,0} \right \rangle$ are not affected by auxiliary bits.

\par If the state sent by Alice is $\left | 0 \right \rangle$, the probability that Eve could guess right Alice's bit under USD measurement is
\begin{equation}
p_{0}^{r}=p_{0,0} \left ( 1-\left \langle E_{0,0} | E_{1,1} \right \rangle \right ) + \frac{1}{2}p_{0,1},
\end{equation}
the probability that Eve could guess wrong Alice's bit is
\begin{equation}
p_{0}^{e}= \frac{1}{2}p_{0,1}.
\end{equation}
The guessing probabilities for $p_{1}^{r}$ and$p_{1}^{e}$ can also be obtained in the same way. So on the $Z$ basis, Eve's quantum USD measurement matrix is
\begin{equation}
p_{z}=A_{z}\begin{bmatrix}
 \frac{p_{0}^{r}}{A_{z}}&\frac{p_{0}^{e}}{A_{z}} \\
\frac{p_{1}^{e}}{A_{z}} & \frac{p_{1}^{r}}{A_{z}}
\end{bmatrix},
\end{equation}
where the normalization constant $A_{z}$ is equal to $4p_{e}-2p_{e}\sin^{2}2\alpha$, and $p_{0}^{r}=p_{1}^{r}$, $p_{0}^{e}=p_{1}^{e}$. Thus under $Z$ base, through collective attack, the Holove quantity between Eve and Alice and Bob $\chi^{z}\left ( E:A,B \right )$ is
\begin{equation}
\chi^{z}\left ( E:AB \right )=A_{z} \left [ 1-H\left ( \frac{p_{0}^{r}}{A_{z}},\frac{p_{0}^{e}}{A_{z}} \right ) \right ],
\end{equation}
where $H\left ( x,y \right )=-x\log _{2}x-y\log_{2}y$. Similarly, under $X$ base, the Holove quantity between Eve and Alice and Bob under collective attack through quantum USD measurement is
\begin{equation}
\chi^{x}\left ( E:AB \right )=A_{x} \left [ 1-H\left ( \frac{p_{+}^{r}}{A_{x}},\frac{p_{+}^{e}}{A_{x}} \right ) \right ],
\end{equation}
where the normalization constant $A_{x}$ is equal to $2p_{e}\left ( 1+ \sin^{2}2\alpha \right )$. Then the total Holove quantity $\chi \left (E: A,B \right )$ is
\begin{equation}
\chi \left (E: AB \right )=\frac{1}{2} \left [ \chi^{z}\left ( E:AB \right )+\chi^{x}\left ( E:AB \right ) \right ].
\end{equation}

\begin{figure}
    \centering
    \includegraphics[width=8cm]{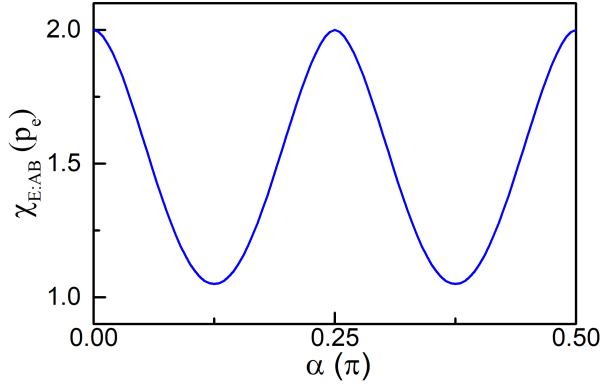}
    \caption{Dependence of the Holove quantity $\chi \left (E;AB \right )$ on $\alpha$. When $\alpha$ is an integer multiple of $\dfrac{\pi}{4}$, $\chi \left ( E;AB \right )$ takes the maximun value $2p_{e}$.}
    \label{fig:my_label}
\end{figure}

\par Fig. 3 shows the simulation result of the dependence of the total Holove quantity $\chi \left (E: AB \right )$ on $\alpha$. We can see that when the value of $\alpha$ is an integer multiple of $\dfrac{\pi}{4}$, $\chi \left (E: AB \right )$ takes the maximum value $2p_{e}$. When the value of $\alpha$ is 0, $\left | E_{0,0} \right \rangle$ and $\left | E_{1,1} \right \rangle$ are located in the direction of $\left | E_{p} \right \rangle$ and $\left | E_{q} \right \rangle$, the amplitudes of $\left | E_{0,1} \right \rangle$ and $\left | E_{1,0} \right \rangle$ are both 0, the BER in $Z$-base is 0, the maximum information Eve can obtain on $Z$ basis is $4p_{e}$. However, for $X$ basis, $\left | E_{+,+} \right \rangle$ and $\left | E_{-,-} \right \rangle$ are both in the direction of $\left | E_{p} \right \rangle+\left | E_{p} \right \rangle$, $\left | E_{+,-} \right \rangle$ and $\left | E_{-,+} \right \rangle$ are in the direction of $\left | E_{p} \right \rangle-\left | E_{p} \right \rangle$, regardless of what quantum state Alice sends, Eve stores the same mixed state, the BER in $X$-base is $2p_{e}$. In this case, the amount of information Eve can obtain is 0. When the value of $\alpha$ is $\dfrac{\pi}{4}$, the situation in $X$-base and $Z$-base under collective attack is completely opposite. In order to balance the BERs on $X$-and $Z$-bases, Eve would be better off doing the following collective attack operation
\begin{equation}
\rho_{ABE}=\frac{1}{2}T\rho_{\phi} \rho_{E} T^{+} +\frac{1}{2}TH\rho_{\phi} \rho_{E}H^{+} T^{+} ,
\end{equation}
where $\rho_{E}=\left | E \right \rangle \left \langle E \right |$ is the state controlled by Eve and $\rho_{\phi}=\left | \phi \right \rangle \left \langle \phi \right |$ is the state sent by Alice, $H$ is a Hadamard operator, which is used to exchange $X$ and $Z$ bases.

\begin{figure}
    \centering
    \includegraphics[width=8cm]{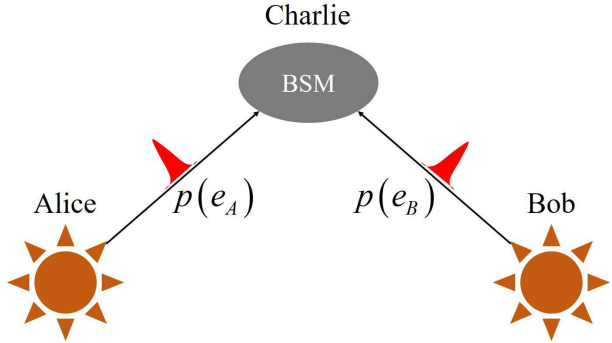}
    \caption{Schematic illustration of Eve's collective attack performed on MDI-QKD. Alice and Bob randomly prepare their states according to the pre-agreed two MUBs, then send them throught two identical fibers to the third party, Charlie, for BSM. The BER $p\left ( e_{A} \right )$, $p\left ( e_{B} \right )$ in the two channels are identical. Eve implements collective attack on the two states with two auxiliary bits, the quantum operations operated on the two channels are strongly correlated.}
    \label{fig:my_label}
\end{figure}

\par The collective attack scheme derived from BB84 protocol can also be applied to MDI-QKD. In MDI-QKD, as shown in Fig. 4, Alice and Bob randomly prepare quantum states according to two sets of MUBs agreed beforehand and send them along two identical optical fibers to the third party, Charlie, for Bell-state measurement (BSM). The BERs on the two channels are $p\left ( e_{A} \right )$ and $p\left ( e_{B} \right )$, respectively. For symmetric MDI-QKD, the BER satisfies $p\left ( e_{A} \right ) \approx p\left ( e_{B} \right )$. When the value of BER is small enough, the total BER $p\left ( e \right )$ of the system is approximately $p\left ( e_{A} \right )+p\left ( e_{B} \right )$. Eve can execute collective attacks on two channels separately, and the total amount of information obtained is $r\leq 2\left ( p\left ( e_{A} \right )+p\left ( e_{B} \right ) \right )$. In MDI-QKD, Eve could execute the following collective attack operation
\begin{equation}
\rho_{ABE}=\frac{1}{2}T\rho_{AB} \rho_{E} T^{+} +\frac{1}{2}TH\rho_{AB} \rho_{E}H^{+} T^{+},
\end{equation}
where $T=T_{A}T_{B}$ is the joint weak measurement operator on channels $A$ and $B$, $H=H_{A}H_{B}$ is the joint Hadamard operator on states sent by Alice and Bob. In order to obtain the information definitely, the weak measurements $T_{A}$ and $T_{B}$ on both sides need to satisfy the condition $\alpha_{A}=\alpha_{B}$, where $\alpha$ is the angle shown in Fig. 2.

\section{Collective attack in entangled QKD protocols}

\par Entangled QKD protocols, such as DI-QKD\cite{acin2007device-independent}, single-photon entanglement based phase-matching-QKD\cite{li2019phase}, and non-entangled QKD protocols, such as BB84-QKD\cite{bennett2011withdrawn}, MDI-QKD\cite{lo2012measurement-device-independent} and twin-field (TF)-QKD\cite{lucamarini2018overcoming}, have many inherent links. Both of these two kinds of QKD protocols need to select two MUBs to prepare quantum states. The final shared key is extracted after announcing the selected base through a trusted public channel. The security proof of non-entangled QKD protocol sometimes uses the method of entanglement distillation\cite{shor2000simple}. The time reversal relationship between MDI-QKD and DI-QKD is satisfied. Therefore, we have reason to believe that there are similarities between collective attacks in non-entangled QKD protocol and those in entangled QKD protocol. Whether in Fock-space based MDI-QKD or wave-space entanglement-based DI-QKD, BSM is an effective method to close the detection loopholes. However, in the conventional QKD protocol based on two-photon entanglement, detection loopholes is one of the main factors limiting its practicability. Because the collective attack is a channel attack scheme, we temporarily ignore the information leak caused by the detection-side loopholes.

\begin{figure}
    \centering
    \includegraphics[width=8cm]{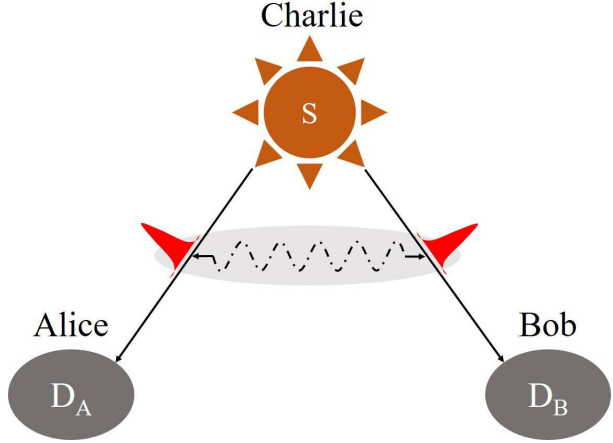}
    \caption{Schematic illustration of Eve's collective attack on DI-QKD. The third party, Charlie, prepares a Bell state and sends the two parts through two identical channels to Alice and Bob. Eve implements collective attack on the two entangled state with two auxiliary bits.}
    \label{fig:my_label}
\end{figure}

\par Fig. 5 shows a brief experimental schematic of DI-QKD. The third party, Charlie, who controls the entanglement resources, sends the two parts of the entanglement to Alice and Bob respectively over two identical channels. Since a Bell state is symmetrical or anti-symmetrical in conjugate spaces, Alice and Bob can select two sets of MUBs (or conjugate measurement bases) to measure the quantum states they receive. After announcing the measurement basis publicly, they randomly select some measurement data to construct Bell-inequality to infer the degree of entanglement and BER of the system. Like the non-entangled QKD protocol, the bit error rate of the system is Eve's operating space for collective attacks.

\par Suppose that in an entangled QKD protocol, a third party, Charlie, prepares the Bell state $\left | \Phi_{AB}^{+}\right \rangle_{Z}=\frac{\sqrt{2}}{2}\left [ \left | 0 \right \rangle_{A}\left | 0 \right \rangle_{B} +\left | 1 \right \rangle_{A}\left | 1 \right \rangle_{B} \right ]$. Alice and Bob choose measurement operators $\sigma_{Z}$ and $\sigma_{X}$ in $Z$ basis and $\frac{\sqrt{2}}{2}\left ( \sigma_{Z}+ \sigma_{X} \right )$ and $\frac{\sqrt{2}}{2}\left ( \sigma_{Z}-\sigma_{X} \right )$ in $X$ basis. Eve attaches his state $\left | E_{A} \right \rangle \left | E_{B} \right \rangle$ to each of the Bell state $\left | \Phi_{AB}^{+} \right \rangle$ independently, then performs a weak measurement operation on the joint state to entangle his state with the joint system of Alice and Bob. In a worse case, Charlie is Eve's puppet, and Eve can instruct Charlie to prepare three-body quantum states as he wishes. These two versions of collective attack are equivalent. In collective attack scheme, Eve's controllable range includes the light source and the two channels, while Alice and Bob's detection sides are isolated from Eve. Alice and Bob extract the key on $Z$-basis and $Z$-basis from the same Bell state, the transformation relations of Bell states in X and Z bases can be expressed as follows
\begin{equation}
\begin{split}
\left | \Phi_{AB}^{+}\right \rangle_{Z} = \left | \Phi_{AB}^{+}\right \rangle_{X}, & \left | \Phi_{AB}^{-}\right \rangle_{Z} = \left | \Psi_{AB}^{+}\right \rangle_{X},\\
\left | \Psi_{AB}^{+}\right \rangle_{Z} = \left | \Phi_{AB}^{-}\right \rangle_{X}, & \left | \Psi_{AB}^{-}\right \rangle_{Z} = -\left | \Psi_{AB}^{+}\right \rangle_{X}.
\end{split}
\end{equation}
By substituting Eqs. (5) and (8) into triplet state $\left | \Phi_{AB}^{+}\right \rangle_{Z}$, we obtain collective attacks on entangled QKD under $Z$- and $X$-base representations
\begin{equation}
\begin{split}
T_{AB} \left | \Phi_{AB}^{+} \right \rangle_{Z} \left | E_{A} \right \rangle \left | E_{B} \right \rangle  &=\frac{\sqrt{2}}{2}  [ \left | E_{00} \right \rangle_{AB}\left | 0 \right \rangle_{A}\left | 0 \right \rangle_{B}
 +\left | E_{11} \right \rangle_{AB}\left | 1 \right \rangle_{A}\left | 1 \right \rangle_{B}\\
&+\left | E_{01} \right \rangle_{AB}\left | 0 \right \rangle_{A}\left | 1 \right \rangle_{B}
+\left | E_{10} \right \rangle_{AB}\left | 1 \right \rangle_{A}\left | 0 \right \rangle_{B}],\\
T_{AB} \left | \Phi_{AB}^{+} \right \rangle_{X} \left | E_{A} \right \rangle \left | E_{B} \right \rangle  &=\frac{\sqrt{2}}{2}  [ \left | E_{++} \right \rangle_{AB}\left | + \right \rangle_{A}\left | + \right \rangle_{B}
+\left | E_{--} \right \rangle_{AB} \left | - \right \rangle_{A}\left | - \right \rangle_{B}\\
&+ \left | E_{+-} \right \rangle_{AB} \left | + \right \rangle_{A}\left | - \right \rangle_{B}
+ \left | E_{-+} \right \rangle_{AB} \left | - \right \rangle_{A}\left | + \right \rangle_{B}],
\end{split}
\end{equation}
where the states in Eve's hands have the following expressions
\begin{equation}
\begin{split}
\left | E_{00} \right \rangle_{AB}&=a^{2}\left | E_{p} \right \rangle_{A}\left | E_{p} \right \rangle_{B}+b^{2}\left | E_{q} \right \rangle_{A}\left | E_{q} \right \rangle_{B},\\
\left | E_{11} \right \rangle_{AB}&=b^{2}\left | E_{p} \right \rangle_{A}\left | E_{p} \right \rangle_{B}+a^{2}\left | E_{q} \right \rangle_{A}\left | E_{q} \right \rangle_{B},\\
\left | E_{01} \right \rangle_{AB}&=\left | E_{10} \right \rangle_{AB}=ab\left (\left | E_{p} \right \rangle_{A}\left | E_{p} \right \rangle_{B}-\left | E_{q} \right \rangle_{A}\left | E_{q} \right \rangle_{B}\right ),\\
\left | E_{++} \right \rangle_{AB}&=\frac{1}{2}\left (\left ( a+b \right )^{2}\left | E_{p} \right \rangle_{A}\left | E_{p} \right \rangle_{B}+\left ( a-b \right )^{2}\left | E_{q} \right \rangle_{A}\left | E_{q} \right \rangle_{B}\right ),\\
\left | E_{+-} \right \rangle_{AB}&=\frac{1}{2}\left (\left ( a-b \right )^{2}\left | E_{p} \right \rangle_{A}\left | E_{p} \right \rangle_{B}+\left ( a+b \right )^{2}\left | E_{q} \right \rangle_{A}\left | E_{q} \right \rangle_{B}\right ),\\
\left | E_{-+} \right \rangle_{AB}&=\left | E_{--} \right \rangle_{AB}=\frac{1}{2}\left (a^{2}-b^{2}\right )\left (\left | E_{p} \right \rangle_{A}\left | E_{p} \right \rangle_{B}-\left | E_{q} \right \rangle_{A}\left | E_{q} \right \rangle_{B}\right ).
\end{split}
\end{equation}
From Eqs. (24) we can see that the quantum states in Eve's hands in $Z$ and $X$ bases are the same as those in Eve's hands, in Eqs. (6) and (9), under collective attack in non-entangled QKD. $\left | E_{00} \right \rangle_{AB}, \left | E_{11} \right \rangle_{AB}$ and $\left | E_{11} \right \rangle_{AB}$ are symmetrically located on each side of $\left | E_{p} \right \rangle_{A}\left | E_{p} \right \rangle_{B}+\left | E_{q} \right \rangle_{A}\left | E_{q} \right \rangle_{B}$, $\left | E_{01} \right \rangle_{AB},\left | E_{10} \right \rangle_{AB}$ and $\left | E_{+-} \right \rangle_{AB},\left | E_{-+} \right \rangle_{AB}$ are in the direction of $\left | E_{p} \right \rangle_{A}\left | E_{p} \right \rangle_{B}-\left | E_{q} \right \rangle_{A}\left | E_{q} \right \rangle_{B}$. Thus the same quantum USD measurements for collective attacks can be applied here. Just like the collective attack scheme in BB84 protocol, Eve can steal the maximum amount of information when $\alpha$ is an integer multiple of $\dfrac{\pi}{4}$.

\par Eve operates on Bell state $\left | \Phi_{AB}^{+} \right \rangle_{Z}$ using a joint operator consisting of a weak measurement operator and a reverse rotation operator for $\alpha=0$,
\begin{equation}
T_{AB} \left | \Phi_{AB}^{+} \right \rangle_{Z} \left | E_{A} \right \rangle \left | E_{B} \right \rangle =\frac{\sqrt{2}}{2} \left [ \left | 0 \right \rangle_{A}\left | 0 \right \rangle_{B} \left | E_{0} \right \rangle_{A} \left | E_{0} \right \rangle_{B} + \left | 1 \right \rangle_{A}\left | 1 \right \rangle_{B} \left | E_{1} \right \rangle_{A} \left | E_{1} \right \rangle_{B} \right ].
\end{equation}
As both $\left | E_{0(1)} \right \rangle_{A}$ and $\left | E_{0(1)} \right \rangle_{B}$ are in Eve's hand, it will be of Eve's convenience to set the joint state $ \left | E_{0} \right \rangle_{A} \left | E_{0} \right \rangle_{B}$ to $\left | E_{0} \right \rangle$, and the joint state $ \left | E_{1} \right \rangle_{A} \left | E_{1} \right \rangle_{B}$ to $\left | E_{1} \right \rangle$. As in Fig. 1, we choose two orthogonal directions $\left | E_{p} \right \rangle+ \left | E_{q} \right \rangle$ and $\left | E_{p} \right \rangle+ \left | E_{q} \right \rangle$ according to $\left | E_{p} \right \rangle$ and $\left | E_{q} \right \rangle$, here we construct a pair of normalized orthogonal bases $\left | E_{\perp} \right \rangle$ and $\left | E_{\parallel} \right \rangle$
\begin{equation}
\begin{split}
\left | E_{\perp} \right \rangle & = \frac{1}{\sqrt{2\left ( 1+ \left \langle E_{0} | E_{1} \right \rangle \right )}} \left [ \left | E_{0} \right \rangle +\left | E_{1} \right \rangle \right ],\\
\left | E_{\parallel} \right \rangle & = \frac{1}{\sqrt{2\left ( 1-\left \langle E_{0} | E_{1} \right \rangle \right )}} \left [ \left | E_{0} \right \rangle -\left | E_{1} \right \rangle \right ].
\end{split}
\end{equation}
So by substituting Eqs. (26) into Eqs. (25), we get the three-body joint state
\begin{equation}
\begin{split}
\left | \Psi_{ABE} \right \rangle & = \frac{\sqrt{2}}{2} \left [ \left | 0 \right \rangle_{A}\left | 0 \right \rangle_{B} \left | E_{0} \right \rangle + \left | 1 \right \rangle_{A}\left | 1 \right \rangle_{B} \left | E_{1} \right \rangle \right ]\\
&=\frac{\sqrt{2}}{2} \left [ \sqrt{1+ \left \langle E_{0} | E_{1} \right \rangle} \left | E_{\perp} \right \rangle\left | \Phi_{AB}^{+}\right \rangle_{Z}+\sqrt{1- \left \langle E_{0} | E_{1} \right \rangle} \left | E_{\parallel} \right \rangle\left | \Phi_{AB}^{-}\right \rangle_{Z} \right ].
\end{split}
\end{equation}

Therefore, according to Eqs. (25, 26, 27), we can see that when Alice and Bob measure in $Z$ basis, the BER of the system is 0, and the maximum amount of information that Eve can obtain based on quantum USD measurement is $1-\left \langle E_{0} | E_{1} \right \rangle$. While in $X$ basis, what Alice and Bob receive is the mixed state
\begin{equation}
\rho_{AB}=\frac{1+ \left \langle E_{0} | E_{1} \right \rangle}{2}  \left | \Phi_{AB}^{+}\right \rangle_{X} \left \langle \Phi_{AB}^{+} \right | +\frac{1- \left \langle E_{0} | E_{1} \right \rangle}{2}  \left | \Psi_{AB}^{+}\right \rangle_{X} \left \langle \Psi_{AB}^{+} \right | .
\end{equation}
In this case, the state of Eve and the joint state of Alice and Bob will be disentangled, Eve will not steal any information, and the BER measured in $X$ basis is $\dfrac{1- \left \langle E_{0} | E_{1} \right \rangle}{2} $. Alice and Bob choose their measurement operations half on the $Z$ basis and half on the $X$ basis, then the total BER of the system is $p_{e}=\dfrac{1- \left \langle E_{0} | E_{1} \right \rangle}{4}$. To balance the BERs on $X$ and $Z$ bases, Eve can perform the following collective attack
\begin{equation}
\rho_{ABE}=\frac{1}{2}\left [ T_{AB}\left | \Phi_{AB}^{+}\right \rangle_{Z} \left \langle \Phi_{AB}^{+} \right | T_{AB}^{+} + H_{AB}T_{AB}\left | \Phi_{AB}^{+}\right \rangle_{Z} \left \langle \Phi_{AB}^{+} \right | T_{AB}^{+} H_{AB}^{+}  \right ],
\end{equation}
where $H_{AB}$ is the joint Hadamard operator used to exchange $X$ and $Z$ bases. It should be noted here that because operators $T_{AB}$ and $H_{AB}$ do not commute, Eve should first perform  the weak measurement operation $T_{AB}$ and then carry out the joint Hadamard transform $H_{AB}$ when conducting collective attacks on $X$-base. Just like the non-entangled QKD protocol, the maximum amount of information that Eve can steal by collective attack on the entangled QKD protocol is $2p_{e}$.

\par Finally, we compare the system key rate and the maximum tolerable BER obtained by security proof based on entanglement purification and security proof against collective attacks in the non-entangled QKD and entangled QKD. The secrete key rate of a QKD is lower bounded by
\begin{equation}
r \geq I\left ( A; B \right ) - \chi \left ( E; AB \right ),
\end{equation}
where $I\left ( A; B \right ) =1-H\left ( p_{e} \right )$ is the mutual information between Alice and Bob, the constraints of the system's tolerable BER is $r>0$. In this discussion, we temporarily ignore all detection loopholes and the impact of post-processing. In entanglement purification protocol, the Holove quantity $\chi \left (E;AB\right )$ between Eve and Alice and Bob in a non-entangled QKD, like BB84- and MDI-QKD, is $H\left (p_{e} \right )$\cite{lo1999unconditional,shor2000simple}, the maximum tolerable BER is $11\%$. While in collective attack scheme, we demonstrate that the maximum value of $\chi \left (E;AB\right )$. When the BER is less than $0.5$, we have $H\left (p_{e} \right )>2p_{e}$, which means that Eve can actually steal less information. The maximum tolerable BER in the security proof against collective attack increases to $17\%$. In entangled QKDs, Eve's eavesdropping is monitored by testing the violation of Bell's inequality. Under the collective attack like Eqs. (29), the Bell function satisfies $S= 2\sqrt{2}\left ( 1-2p_{e} \right )$. Quantum non-localized correlation between Alice and Bob requires $S > 2$. The maximum tolerable BER in a entangled QKD based on the security proof against collective attack is $14.6\%$. Therefore, the upper bound of key rate is obtained by entanglement purification protocol, and the lower bound of key rate is obtained by security against collective attack.

\section{Conclusion}

\par In this paper, we deduce the maximum amount of information Eve can steal through collective attacks for different QKD protocols. We find that in different QKD protocols, including non-entangled and entangled QKDs, Eve's collective attacks scheme based on quantum weak measurement are equivalent. Among these attacks, the maximum amount of information Eve can steal is $2p_{e} $. This proof not only enhances the maximum bit error rate tolerated by the system, but also enhances the maximum key rate that can be extracted. This work not only gives a clear lower bound for the key rate of security proof against all collective-attacks, but also helps us to have a more comprehensive and deeper understanding of collective attacks in different kinds of QKDs.

\acknowledgments
This work is supported by Young fund of Jiangsu Natural Science Foundation of China (SJ216025), National fund incubation project (NY217024), the National Natural Science Foundation of China (No. 61871234), Scientific Research Foundation of Nanjing University of Posts and Telecommunications (NY215034), the open subject of National Laboratory of Solid State Microstructures of Nanjing University (M31021).

\bibliography{reference}
\end{document}